\documentclass[sort&compress,square, numbers,sort,nofootinbib]{revtex4-2}

\usepackage[utf8]{inputenc}
\usepackage{url} 
\usepackage{float}
\usepackage{amsmath, amssymb}
\usepackage{graphicx}
\usepackage{caption}
\usepackage{subcaption}

\newcommand{\liem}{\mathfrak{L}_m}
\newcommand{\lieb}{\mathfrak{L}_{\beta}}
\newcommand{\GRB}{\texttt{GRBoondi}  }

\begin{document}

\title{GRBoondi: A code for evolving Generalized Proca theories on arbitrary backgrounds}
	
\author{Shaun David Brocus Fell}
\email{fell@thphys.uni-heidelberg.de}
\affiliation{Institute for Theoretical Physics, Universit{\"a}t Heidelberg , Philosophenweg 16, 69120 Heidelberg, Germany}

\author{Lavinia Heisenberg}
\email{heisenberg@thphys.uni-heidelberg.de}
\affiliation{Institute for Theoretical Physics, Universit{\"a}t Heidelberg , Philosophenweg 16, 69120 Heidelberg, Germany}

\begin{abstract}
While numerical simulations offer unparalleled precision and robustness in studying complex physical systems, their execution is often hindered by complexity, costliness, and time consumption due to the intricate equations involved. This challenge is already encountered in General Relativity, where non-flat spacetimes exacerbate the computational burden. 
This complexity is further intensified when dealing with additional degrees of freedom.
To address this challenge head-on, we introduce \GRB, a groundbreaking fixed-background numerical relativity code designed to provide a unified interface for numerically solving Generalized Proca theories. \GRB grants users the ability to make arbitrary modifications to the Proca equations of motion on any background, providing a robust and versatile tool for exploring diverse classes of Generalized Proca theories. This letter serves as part of the submission of \GRB to the Journal of Open Source Software. For access to the code, please visit https://github.com/ShaunFell/GRBoondi.git.


\end{abstract}

\maketitle

\section{Introduction}
Generalized Proca theories~\cite{Heisenberg_2019, Heisenberg_2014} provide a rich landscape to search for solutions to deep, fundamental questions, such as the nature of dark energy and dark matter. Moreover, strong gravity regimes, like those surrounding dense, compact astrophysical entities, offer novel avenues to explore fundamental fields. To effectively probe these domains, precise models are imperative for sifting through vast quantities of data. Crafting concrete models for the entirety of Generalized Proca theories represents a formidable endeavor, typically relying on numerical methods. These numerical methods are usually written to be case dependent, especially for a particular background. This makes it difficult to generalize the computational code to account for higher-order couplings or different types of backgrounds. The difficulty is especially amplified when considering the full landscape of Generalized Proca theories and beyond~\cite{Heisenberg:2016eld}. 

Generalized Proca theories represent the most general ghost-free theories for a massive vector field that preserve the second-order equations of motion~\cite{Heisenberg_2019, Heisenberg_2014, BeltranJimenez:2016rff, Allys:2015sht, Tasinato:2014eka}. Being the most general construction, it naturally contains numerous highly non-linear derivative interactions of the Proca field. Its defining Lagrangian can be split into six separate interactions determined by the order of the derivatives of the Proca field. Already at the order of the quadratic Lagrangian, many interesting properties of non-linear Proca theories become manifest~\cite{Aoki_Minamitsuji_2022, Barausse_2022, Clough_2022, Coates_2022, PhysRevD.107.104036, _nl_t_rk_2023}. On the astrophysical side, interest in Generalized Proca theories as dark matter and dark energy candidates has seen recent growth~\cite{DeFelice:2016yws, DeFelice:2016uil, DeFelice:2016cri, Heisenberg:2017hwb, Heisenberg:2017xda, deFelice:2017paw, GENG2021100819, 2023PhRvD.107l3535G, de_Rham_2022}. In the pursuit of investigating this class of theories within strong gravity regimes, numerical simulations emerge as an indispensable tool. Additionally, it becomes imperative to constrain the landscape of allowed interactions and parameters associated with these theories.

\GRB\footnote{Consistent with the GRChombo naming convention, Boondi is a Koori word for a multi-purpose tool, used for hunting and digging.} provides a unified interface for computing the evolution of any Generalized Proca model on an arbitrary background. Given a specific background with known expressions for the metric variables and initial data for the Proca field, \GRB numerically computes the time evolution of the Proca field, given user-specified Generalized Proca equations of motion. While the simultaneous evolution of the metric and matter fields offers the most comprehensive depiction of their temporal progression, there are instances where the density of the Proca field is negligible in comparison to the background curvature. In such cases, fixed background evolution routines serve as an excellent approximation to the complete mutual evolution. A core feature of \GRB is that it simplifies many of the boiler-plate code required to begin a simulation from scratch. The only additions a user needs to input are the initial conditions, the modifications to the equations of motion (EOM), and the background functions. \GRB will automatically compute various diagnostic quantities and plot files. Moreover, \GRB is incredibly modular and modifications to any of the in-built functions is effortless.

The core challenge in numerical computations of Generalized Proca theories lies in formulating the EOM's in a manner conducive to time evolution-based solvers. This is usually done by performing a (3+1)-decomposition of the equations of motion and solving for the evolution operator acting on the state variables. For configuration space variables $\vec{Q}$, this amounts to solving the EOM's for the quantities $\liem \vec{Q}$, where $\liem$ is the Lie derivative along the hypersurface evolution vector, henceforth called the evolution operator. Once this quantity has been solved for, it can be split into a time evolution piece and a hypersurface-transverse piece, $\liem \vec{Q} = \frac{1}{\alpha} \left( \dot{\vec{Q}} - \lieb \vec{Q} \right )$, where $\alpha$ is the lapse function and $\lieb$ is the Lie derivative along the shift vector. Once the evolution operator has been solved for, the evolution equation can be easily recast in the form 
\begin{equation} \label{eq:qdot}
\dot{\vec{Q}} = F(\vec{Q}, \vec{B})\;,
\end{equation}
where $F$ is some known function of all the other configuration space variables and the background variables $\vec{B}$. For known expressions for the background variables $\vec{B}$ and specifying $\vec{Q}$ to be the Proca (3+1)-variables,  \GRB numerically solves Eq.(\ref{eq:qdot}) using standard finite-differencing techniques. For more specifics on the (3+1)-splitting, see for example~\cite{Zilhão_Witek_Cardoso_2015, Gourgoulhon2012}.

The fixed background approximation provides a huge speedup compared to the full evolution, since only a small fraction of the variables actually require a time integration. Not only do simulations now require far less resources, the speed up can be up to hundreds of times faster. \GRB is based on the publicly available numerical relativity (NR) code GRChombo \cite{Andrade2021, Clough2015sqa}, which itself is based on the open source adaptive mesh refinement (AMR)-based differential equation solver Chombo \cite{Adams:2015kgr}. Some minor pieces of \GRB are also based off the recently released open source software GRDzhadzha \cite{Aurrekoetxea2024}, which is also based off of GRChombo.

\section{Statement of Need}

In practice, any NR software library can be used to evolve Generalized Proca theories. Prominent examples of NR codes encompass the extensive Einstein toolkit \cite{EinsteinToolkit:2023_11} and its related Cactus framework \cite{Goodale:2002a}, Kranc \cite{Kranc:web}, LEAN \cite{Sperhake:2006cy}, and Canuda \cite{Canuda:zenodo}. Other libraries worth mentioning are the non-public BAM \cite{Bruegmann:2006ulg}, AMSS-NCKU \cite{Galaviz:2010mx}, PAMR \cite{East:2011aa} and HAD \cite{Neilsen:2007ua}. The non-exhaustive list can be expanded with
SPeC \cite{Pfeiffer:2002wt}, which is a pseudo-spectral code that uses generalized harmonic coordinates. Similarly, SpECTRE \cite{deppe_nils_2021_4734670, Kidder:2016hev, Cao:2018vhw} uses the Galerkin methods. NRPy \cite{Ruchlin:2017com} is a python library that aims for use on non-high performance computing clusters. There are also cosmological simulation codes like CosmoGRaPH \cite{Mertens:2015ttp} and GRAMSES \cite{Barrera-Hinojosa:2019mzo}. Simflowny \cite{Palenzuela:2018sly} is a magneto-hydrodynamic simulation software. GRAthena++ \cite{Daszuta:2021ecf} uses the oct-tree AMR for maximum scaling. GRDzhadzha \cite{Aurrekoetxea2024} is another fixed background code, based off the GRChombo \cite{Andrade2021} framework.

This extensive list underscores the abundance of NR libraries at one's disposal. However, none of them provide a tailored unified interface for studying the vast landscape of gravity theories, like Generalized Proca. Using any of the numerous frameworks requires significant work in order to evolve even a single Generalized Proca theory. In order to gauge the simplicity of using \GRB, we compare the number of lines-of-code (LOC) needed to setup a standard Proca simulation in GRChombo, GRDzhadzha, and \GRB. GRChombo requires over 1500 lines of code, spread across 17 different files, including source and header files. GRDzhadzha requires over 600 LOC across 14 files, and \GRB requires just 300 LOC across 6 files. These simple metrics provide a decent gauge of the complexity faced by the user for setting up a simple standard Proca simulation. On top of this, \GRB offers catered plotting routines for viewing data, leveraging the highly parallelizable VisIt \cite{visit_dav} analysis tool. In addition, since the metric variables and their derivatives are computed exactly at each grid point, the adaptability of the AMR grid can be focused solely on the matter variables. 

While backreaction is neglected in \GRB, the error incured by this approximation is estimated by computing the norm of the energy-momentum tensor. This estimation takes into account the relativistic nature of the matter field and gauges the error in the evolution at the level of the Einstein equations. 

Since \GRB is very similar to the GRChombo code, simulations performed with \GRB can easily be ported to GRChombo, should full NR simulations be required. This is particularly useful if the backreaction is found to be significant at some point in the simulation. The data from a \GRB simulation can be used as initial data for a GRChombo simulation, potentially yielding much better initial data for the full NR simulation.

\section{Key Features of GRBoondi}

\begin{itemize}
\item \textbf{Ease of use}: A central pillar of \GRB is its ease of use, relative to other NR software. Many of the basic boiler-plate code and complexities are kept within the source code, allowing the user to have only a basic understanding of the code in order to start researching their problem.

\item \textbf{Arbitrary choice of background spacetime}: The main parts of \GRB are extremely modular, allowing for any arbitrary background to be plugged in, even a numerically computed one. This allows for massive versatility in the source code. Users can swap in and out backgrounds with ease. \GRB comes pre-equipped with four background classes ready for use, along with testing suites to verify convergence of each class. These include:
    \begin{itemize}
        \item Minkowski space
        \item Boosted Schwarzschild black hole
        \item Kerr black hole
        \item Kerr-de Sitter black hole
    \end{itemize}

\item \textbf{Arbitrary modifications to base Proca}: \GRB uses specific coding idioms that allow arbitrary modifications to the hard-coded equations of motion. This allows for numerical simulations of any Generalized Proca theory. The base model is the standard electromagnetic model, subject to the usual Proca constraint. Any Generalized Proca theory can be added on top of this by simply adding the additional pieces from the Generalized Proca Lagrangian. Examples of simulating standard Proca and a simple non-linear model \cite{coates2022intrinsic, Clough_2022, _nl_t_rk_2023} are included in the codes repository.

\item \textbf{Accuracy}: The metric values and their derivatives are computed exactly at each point, while the matter variables are evolved using 4th-order Runge-Kutta time integration and their spatial derivatives calculated using the same finite difference stencils as GRChombo (currently, 4th- and 6th-order stencils are available).

\item \textbf{Various boundary conditions}: Since \GRB inherits from GRChombo, all the boundary conditions in GRChombo are also straight-forwardly applicable to \GRB. \GRB thus supports:
    \begin{itemize}
         \item Periodic boundary conditions
         \item Sommerfeld out-going wave conditions
         \item Static boundary conditions
         \item Extrapolation at linear and zeroth order
        \item Reflective symmetry (e.g. simulating only the top-half of the computational box due to the background symmetry)
    \end{itemize}

\item \textbf{Checkpointing}: Since \GRB inherits from Chombo, it inherits the checkpointing feature. Simulations can be restarted from a checkpoint file saved during the previous run, allowing for arbitrarily long simulations, even on restricted computing clusters. Moreover, should a simulation fail for any reason, it can be restarted using a previous checkpoint.

\item \textbf{Diagnostics}: \GRB comes equipped with many different diagnostic quantities that can be calculated during the course of a simulation and additional tools for users to calculate their own quantities.
    \begin{itemize}
        \item \emph{Computation of conserved quantities}. Computation of conserved quantities, such as energy and angular momentum densities and their associated fluxes across a surface can be toggled. \GRB can also calculate the energy-momentum tensor trace and its square, for studying the importance of backreaction. Some of the included diagnostic quantities are the square of the Proca field,  conserved energy density, Eulerian energy density, conserved angular momentum, conserved energy density flux, conserved angular momentum flux, trace of the energy-momentum tensor, and square of the energy-momentum tensor. The conserved energy and angular momentum correspond to approximate time and rotational killing vectors. These can be disabled if the spacetime does not possess these.
        \item  \emph{Excision of diagnostics in user-specified regions}. Currently, only spherical excision is implemented, where the user specifies the minimum and maximum radius for the diagnostic quantities. The user can specify which quantities they wish to excise as well, as specified in a parameter file. Generalization to more complicated diagnostic regions is case dependent and left up to the user.
        \item  \emph{Integration of quantities across the computational grid}. The user can specify which quantities they wish to integrate.
        \item  \emph{Integration of quantities across spherical surfaces}. The user can specify which quantities they wish to integrate over a spherical surface and the radii of those surfaces as well.
    \end{itemize}

\item \textbf{Tailored post-processing tools}: \GRB comes equipped with custom-built post-processing scripts, leveraging VisIt's python interface. Any of the diagnostic quantities computed during the evolution can be plotted using these tools. It also contains a simple integral plotter, for quickly visualizing the various integrals computed during a simulation, leveraging python's Matplotlib package \cite{Hunter:2007}.
\end{itemize}

\begin{figure}
\centering
\begin{subfigure}{.49\linewidth}
    \centering
    \includegraphics[width=\linewidth]{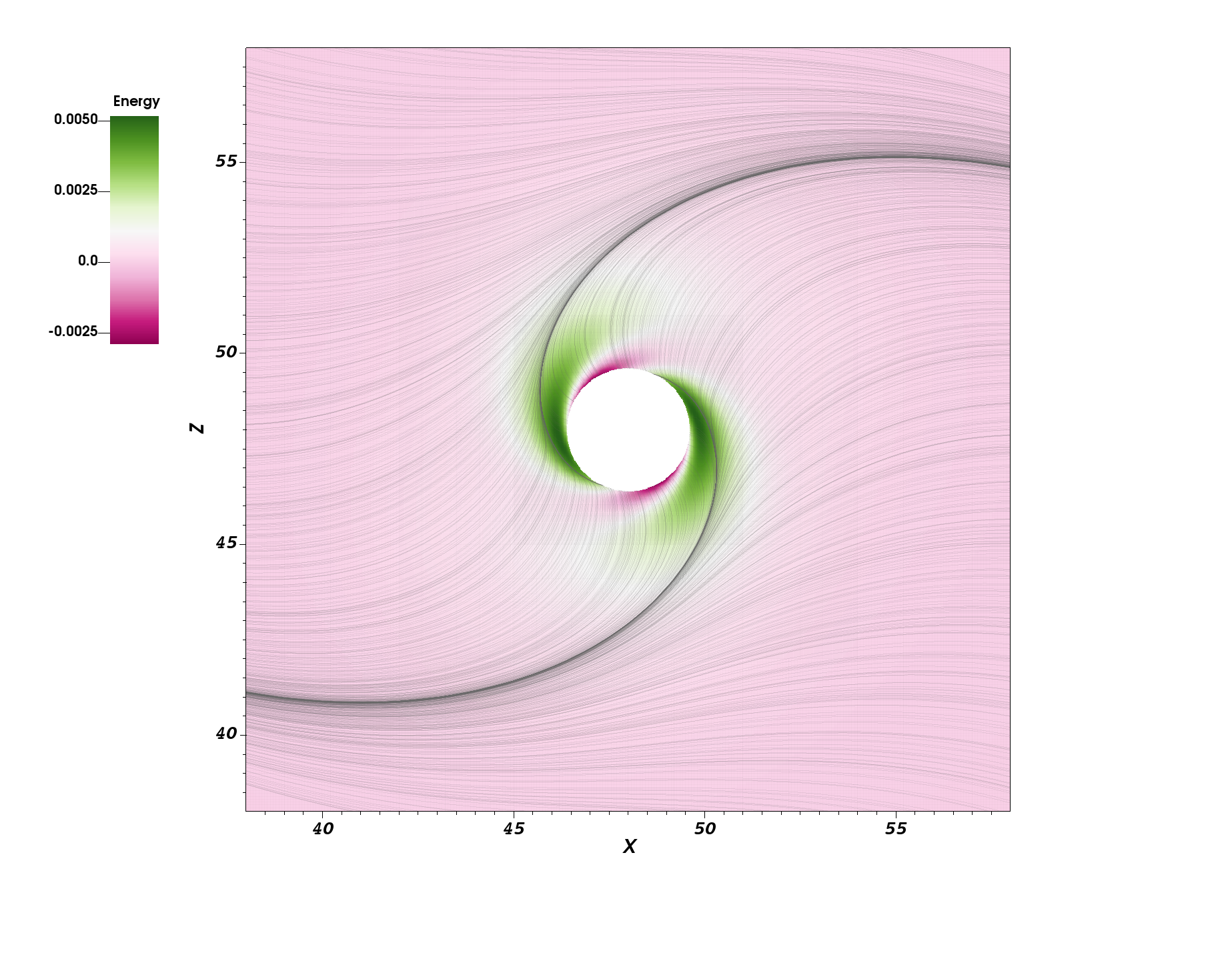}
    \label{fig:fig1}
\end{subfigure}
\begin{subfigure}{.49\linewidth}
    \centering
    \includegraphics[width=\linewidth]{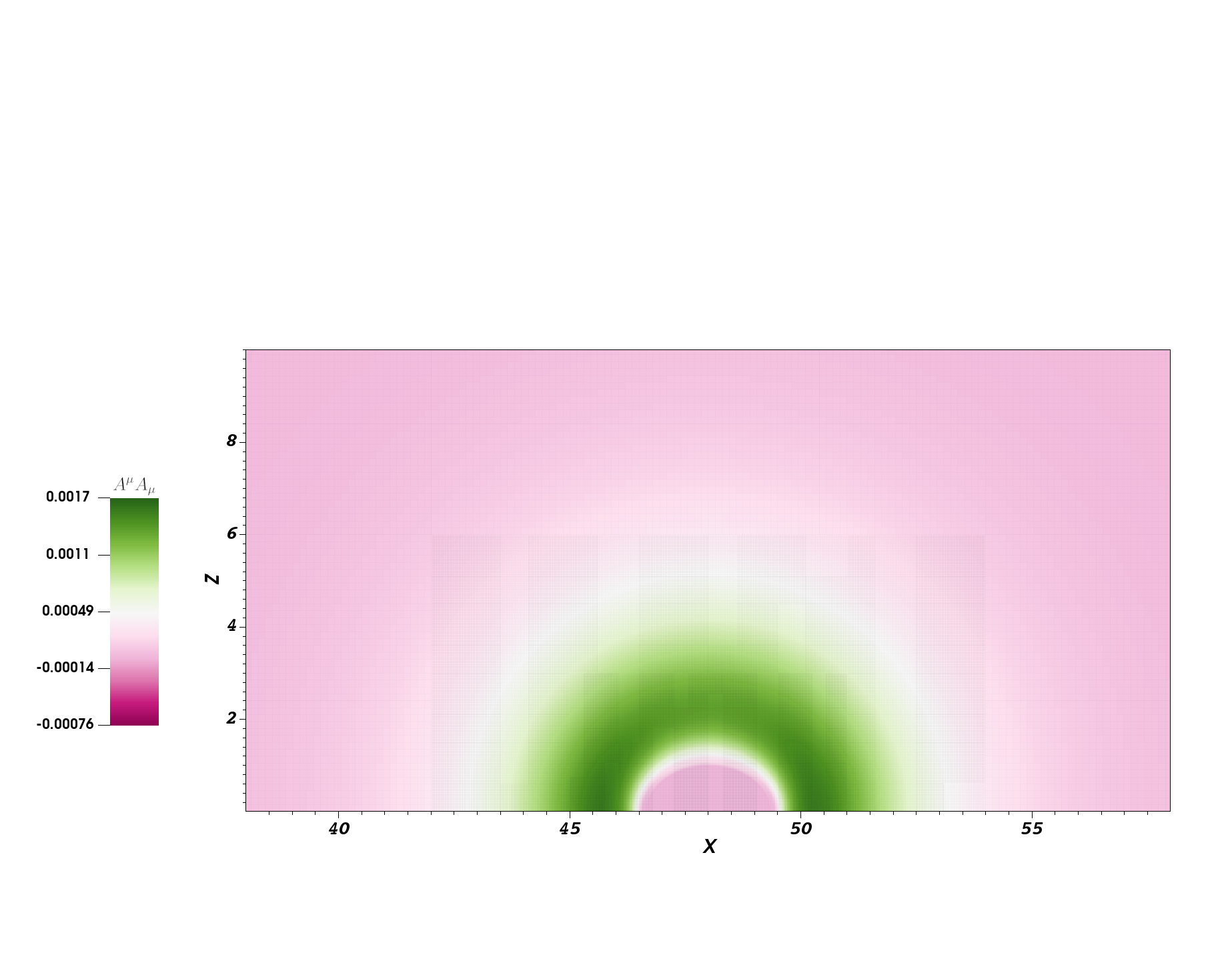}
    \label{fig:fig2}
\end{subfigure}
\caption{Some examples of the usage of \GRB. The left image is the energy density of a Proca cloud superradiantly excited around a rapidly spinning black hole, superimposed with streamlines of the spatial Proca vector field. The right image is a plot in the xz-plane of the square of the Proca 4-vector in the same background. The spin axis of the black hole is orientated along the z-axis. Superimposed on both images is a slice of the computational grid, clearly showing the refinement hierarchy.}
\end{figure}

\section{Summary}
\GRB provides a unified interface for computing the evolution of any Generalized Proca model on an arbitrary background spacetime. It is a powerful tool that opens up many avenues for possible research studies, including cosmological, superradiant, modified gravity, and more. Moreover, having GRChombo as a foundation allows its solutions to be easily ported as initial conditions for a full Einstein+Generalized Proca evolution. We intend to explore this avenue in a future follow-up project. Its modularity allows for great user freedom, thus allowing \GRB to simulate the full landscape of Generalized Proca theories. Its in-built post-processing routines allows for quick visualization of simulation data, both during and after computations.  \GRB will thus be a vital tool in studying Generalized Proca theories, opening the doors to probing astrophysically relevant models of dark matter and dark energy.

\section{Acknowledgements}
LH is supported by funding from the European Research Council (ERC) under the European Unions Horizon 2020 research and innovation programme grant agreement No 801781. LH further acknowledges
support from the Deutsche Forschungsgemeinschaft (DFG, German Research Foundation) under Germany’s Excellence Strategy EXC 2181/1 - 390900948 (the Heidelberg STRUCTURES Excellence Cluster).

SF is indebted to Katy Clough for her stimulating discussions and extensive input to the code. Katy was an invaluable source of aid for the nuances of NR simulations and some of the finer details of GRChombo. 

The simulations performed as part of this release were carried out on the Baden-Württemberg high-performance computing cluster. The authors acknowledge support by the state of Baden-Württemberg through bwHPC.

\bibliographystyle{apsrev4-1}
	\bibliography{paper.bib}

\end{document}